\def \PR {{ Phys. Rev.} }
\def \PL {{ Phys. Lett.} }

\def \NP {{ Nucl. Phys.} }
\def \PRL {{ Phys. Rev. Lett. }}
\def \bc {\begin{center}}
\def \ec {\end{center}}

\def \bfr {\begin{flushright}}
\def \efr {\end{flushright}}

\def \ba {\begin{array}}
\def \ea {\end{array}}

\def \bea {\begin{eqnarray}}
\def \eea {\end{eqnarray}}

\def \be {\begin{equation}}
\def \ee {\end{equation}}
\def\ni{\noindent}
\def\nn{\nonumber}
\def\p{\partial}
\def\f{\frac}
\def\l[{\left[}
\def\r]{\right]}
\def\TG{\tilde{G}}
\def\tg{\tilde{g}}
\def\TT{\tilde{T}}

\def\um{\frac{1}{2}}

\newcommand{\xr}[1]{ {\tilde{X}}^{R}_{#1} }

\documentstyle[12pt,a4]{article}

\begin{document}

\begin{flushright}
{\large SWAT-99/226}
\end{flushright}

\begin{center} 
{\large {\bf CENTRAL EXTENSIONS AND QUANTUM PHYSICS}}
\footnote{Work partially supported by the DGICYT.}
\end{center}
\bigskip
\bigskip
\centerline{ {\sc M. Calixto$^{1,3}$\footnote{E-mail: pymc@swansea.ac.uk / 
calixto@ugr.es} and  
 V. Aldaya$^{2,3}$\footnote{E-mail: valdaya@iaa.es}} }
\bigskip

\begin{enumerate}
\item {Department of Physics, University of Wales Swansea, Singleton Park, 
Swansea, SA2 8PP, U.K.}
\item {Instituto de Astrof\'{\i}sica de Andaluc\'{\i}a, Apartado Postal 3004,
18080 Granada, Spain.}
\item  {Instituto Carlos I de F\'\i sica Te\'orica y Computacional, Facultad
de Ciencias, Universidad de Granada, Campus de Fuentenueva, 
Granada 18002, Spain.} 
\end{enumerate}

\bigskip

\begin{center}
{\bf Abstract}
\end{center}
\small          

\begin{list}{}{\setlength{\leftmargin}{3pc}\setlength{\rightmargin}{3pc}}
\item The unitary implementation of a symmetry group $G$ of a 
classical system in the corresponding quantum theory 
entails unavoidable deformations $\TG$ of $G$, namely, 
central extensions by the typical phase invariance group $U(1)$.  The 
appearance of central charges in the corresponding 
Lie-algebra quantum commutators, as a consequence of 
non-trivial responses of the phase of the  
wave function under symmetry transformations, lead to a 
quantum generation of extra degrees of freedom with regard to the classical 
counterpart. In particular, 
symmetries of the Hall effect, Yang-Mills and conformally invariant 
classical field theories 
are affected when passing to the quantum realm. 

\end{list}

\vskip 1cm
\section{Introduction\label{intro}}

The importance of {\it phase invariance} in 
Quantum Mechanics and its physical implications has been 
widely discussed in the study of, for example, geometric phases 
(see \cite{gphases} for a collection of papers). Fundamental mathematical 
structures and objects like {\it fibre bundles} and {\it holonomies}, 
and  important physical concepts like {\it Berry's phase}, are 
directly attached to this relevant invariance of quantum phenomena. Also, 
the quantization of some physical magnitudes and the 
stability of some topological quantum numbers can be directly 
attributed to the cyclic (compact) character of the quantum mechanical 
phase $\zeta=e^{i\alpha}$. 

However, despite the relevance of phase invariance, the 
standard (canonical) approach to Quantum Theory does not stress enough 
its central role in the quantization procedure. 
Concepts like {\it anomaly} or {\it no-go theorem}, which usually 
make reference to failures associated with the quantization of a classical 
system, have been serious obstacles for the traditional 
quantization methods, even though {\it they} sometimes prove to be essential 
ingredients of the corresponding quantum system. Anomalies, for example, 
which manifest themselves  through the appearance of 
central terms at the right-hand 
side of some commutators, which their classical counterpart (Poisson bracket) 
does not possess, 
prove to have an important physical meaning in connection with  
the {\it spontaneous breakdown} of some underlying symmetry. This central 
term contribution traces back to a non-trivial transformation of the phase 
of the wave function under the consecutive action of two different symmetry 
operations which, nevertheless, leave the corresponding 
classical equations of motion unchanged (see below in Sec. 
\ref{conformesec} in connection with 
the breakdown in the unitary implementation of the conformal symmetry 
for massless fields).

All of these failures in quantization should disappear if one gives 
up the idea of ``quantizing classical systems''. Also, 
any reasonable attempt to reconcile Gravity 
with Quantum Mechanics should 
pass through an {\it inherently} quantum approach to the formulation 
of the quantum theory of the gravitational field. Suitable 
candidates for an {\it intrinsic}  formulation of Quantum Mechanics 
(without any reference to an underlying 
classical system)
should incorporate tools and concepts of Algebra, Geometry and Topology 
to take into account both local and global aspects of Quantum Theory. 
In fact, physical systems possessing a gauge symmetry share and exhibit a 
{\it principal bundle} internal structure where the phase 
invariance plays a ``central role'' in, for example, 
determining the structure of constraints (first- and/or second-class). 
Symmetry is another essential component of Quantum Mechanics; after all, 
any consistent (non-perturbative) 
quantization is mostly a unitary irreducible representation of a 
suitable (Lie, Poisson) algebra. All these previous 
requirements advocate a group structure as a firm candidate 
to define a quantum system. Central extensions $\TG$ of a group $G$ by 
$U(1)$ (the group of phase invariance) have a natural 
fibre bundle structure  $\TG\rightarrow \TG/U(1)$, where 
$U(1)$ is the {\it structure subgroup} of $\TG$ \cite{GAQ}. Other fibrations 
$\TG\rightarrow \TG/\TT$ of $\TG$ by $\TT\sim T\times U(1)$ 
correspond to a constrained quantum system \cite{Ramirez} 
with ``gauge'' group $T$ 
and, as already mentioned, the structure of $\TT$ (as a central extension 
of $T$ by $U(1)$) determines the nature of constraints: first- and/or 
second-class (see later). The group law $\tg''=\tg'*\tg$ 
for $\TG$ can be written in general 
form as:
\be 
\tg''=(g'';\zeta'')=(g'*g;\zeta'\zeta e^{\frac{i}{\hbar}\xi(g',g)}),\,\,\,
g,g',g''\in G;\,\,\, \zeta'',\zeta',\zeta\in U(1),
\ee
where $\xi:G\times G\rightarrow \Re$ is a two-cocycle, which verifies 
the general property:
\be
\xi(g_2,g_1)+\xi(g_2*g_1,g_3)=\xi(g_2,g_1*g_3)+\xi(g_1,g_3)
\;\;,\forall g_i\in G\,, 
\ee
\ni and the constant $\hbar$ (namely, the Planck constant) 
is intended to kill any possible dimension of $\xi$ (namely, action 
dimension). In the general theory of central extensions \cite{Extensiones},  
 two-cocycles are said to be equivalent if they differ by a coboundary, 
i.e. a two-cocycle which can be written in the form  $\xi(g',g)=\eta(g'*g)-
\eta(g')-\eta(g)$, where $\eta(g)$ is 
called the generating function of the coboundary (from now on we shall 
omit the prefix ``two'' when referring to  two-cocycles). 
However, although cocycles 
differing by a coboundary lead to equivalent central extensions as such, 
there are some coboundaries which provide a non-trivial connection
\be 
\Theta=\left.\f{\p}{\p g^j}\xi(g',g)\right|_{g'=g^{-1}}dg^j
-i\hbar\zeta^{-1}d\zeta\,,\label{thetagen}
\ee
on the 
fibre bundle $\TG$ and Lie-algebra structure constants different from those  
of the direct product $G\times U(1)$. These are generated by a function 
$\eta$ with a non-trivial gradient at the 
identity $\left.d\eta(g)\right|_{g=e}=
\left.\frac{\partial\eta(g)}{\partial g^j}\right|_{g=e}d g^j\not=0$, 
and can be divided into pseudo-cohomology equivalence subclasses: 
two {\it pseudo-cocycles} are equivalent if 
they differ by a coboundary generated 
by a function with trivial gradient at the identity 
\cite{Saletan,Pseudoco,Marmo2}. Pseudo-cohomology plays 
an important role in the theory of 
finite-dimensional semi-simple groups, as they have trivial cohomology. For 
them, pseudo-cohomology classes are associated with coadjoint orbits 
\cite{Marmo2}. We shall have the opportunity of showing how, in fact, the 
introduction of coboundaries in some physical systems alters the 
corresponding quantum theory.

Wave functions $\psi$ are defined in this 
group-framework as complex functions 
on $\TG$, $\psi:\TG\rightarrow C$, which verify the {\it $\TT$-equivariance 
condition}
\be
L_{\tg_t}\psi(\tg)\equiv\psi(\tg_t *\tg)=
D_{\TT}^{(\epsilon)}(\tg_t)\psi(\tg)\,,\;\;\forall 
\tg_t\in \TT,\,\,\forall
\tg\in \TG\,,\label{tequiv}
\ee
where $D_{\TT}^{(\epsilon)}$ symbolizes a specific 
representation $D$ of $\TT$ with 
$\epsilon$-index (in particular, the $\epsilon=\vartheta$-angle 
\cite{Jackiwtheta} of non-Abelian gauge theories; see below) and 
generalizes the typical phase invariance of quantum mechanics,  
$D_{\TT}^{(\epsilon)}(\zeta)=\zeta,\,\,\,\forall \zeta\in U(1)\subset\TT$, 
when extra {\it constraints} are considered in the theory. The left-action 
\be
L_{\tg'}\psi(\tg)\equiv\psi(\tg' *\tg),\,\,\tg',\tg\in\TG \label{repre}
\ee
defines a reducible (in 
general) representation of $\TG$ on the linear space of $\TT$-equivariant 
wave functions (\ref{tequiv}). The reduction is achieved by means of those 
right restrictions on wave functions 
\be
R_{\tg_p}\psi(\tg)\equiv\psi(\tg *\tg_p)=\psi(\tg),\,\, \forall 
\tg_p\in G_p,\,\forall \tg\in \TG\label{pola}
\ee
(which commute with the left action)  
compatible with the $U(1)$-equivariant condition 
$D_{\TT}^{(\epsilon)}(\zeta)=\zeta$; the right restrictions (\ref{pola}) 
generalize the notion of {\it polarization conditions} 
of Geometric Quantization (see \cite{GAQ} and references therein), 
the Schr\"odinger equation being a particular case of polarization condition 
in the Group Approach to Quantization (GAQ) framework  \cite{GAQ}. 
The subgroup $G_p$ 
is called the {\it polarization subgroup} (see \cite{chorri,Marmo2} 
for anomalous cases) and contains non-symplectic (without dynamical content) 
coordinates like time, rotations, etc, and half of the symplectic ones 
(either `positions' or `momenta'). The {\it characteristic subalgebra} 
${\cal G}_c$ of left-invariant vector fields (l.i.v.f.) related to 
non-symplectic coordinates admits an algebraic characterization as 
follows:
\be
{\cal G}_c={\rm Ker}\Theta\cap{\rm Ker}d\Theta=
\{{\rm l.i.v.f.} \,\tilde{X}\,/ \, \Theta(\tilde{X})=0,\,
d\Theta(\tilde{X})=0\}\,.\label{cara}
\ee
The representation (\ref{repre}) is also unitary with 
respect to the scalar product
\be
\langle \psi | \psi'\rangle=\int_{\TG}{\mu(\tg)\psi^*(\tg)\psi'(\tg)}
\ee
where $\mu(\tg)=\theta^L_1\wedge\stackrel{{\rm dim}(\TG)}{\dots}\wedge
\theta^L_n$ is the natural left-invariant measure of $\TG$ (exterior product 
of left-invariant one-forms). 

The classical limit of the theory corresponds to the replacement 
$U(1)\leftrightarrow \Re$; that is, to a central extension $\bar{G}$
of $G$ by the additive group $\Re$ with group law 
\be
\bar{g}''=(g'';r'')=(g'*g; r'+r+\xi(g',g)),\,\,\,r,r'\in\Re\,,
\ee
the classical analogue of $\psi$ being Hamilton's 
principal function  when the $U(1)$-equi-\break variance condition 
%$D_{\TT}^{(\epsilon)}(\zeta)=\zeta$ in 
(\ref{tequiv}) is replaced by 
an $\Re$-equivariance condition 
%$D_{\bar{T}}^{(\epsilon)}(r)=0$ 
\be
L_{r}\psi(\bar{g})=\psi(\bar{g})+r\,,\;\;\forall 
r\in \Re,\,\,\forall
\bar{g}\in \bar{G}\,,
\ee

\ni (see \cite{GAQ} for more details). Note that 
the constant $\hbar$ is dispensable in the classical limit $\bar{G}$ of 
$\TG$.

In what follows, we are going to make use of some 
relevant examples where symmetry determines the 
quantum physical system, trying to capture much of the 
GAQ flavour. Let us begin working out a simple, although general, example of a quantizing group 
$\TG$ which eventually applies to a diversity of physical systems, 
for example, the quantum Hall effect and quantum Yang-Mills theories. 

\section{Generalized Conformal Symmetry and Extended Objects from the Free 
Particle\label{extended}}
 
The simplest  
quantum commutation relations are given by the Heisemberg-Weyl commutators:
\be
\left[ \hat{q}_j, \hat{p}_k\right]=i\hbar \delta_{jk}\hat{I}\label{HW}
\ee
which means that, after  successive translations $q'$ and $p'$ 
of the position $q$ and the 
momentum $p$, the wave function gains a non-trivial phase factor 
$\zeta=e^{\frac{i}{2\hbar}\sum_{j=1}^d{(q_j'p_j-p'_jq_j)}}$ which 
can be read from 
the Heisemberg-Weyl group law: 
\be\ba{rcl}
\vec{V}''&=&\vec{V}'+\vec{V} \\ 
\zeta''&=&\zeta'\zeta e^{\frac{i}{2\hbar} \vec{V}'^t W\vec{V}}\ea,\label{HWlaw}
\ee
where we have denoted $\vec{V}=\left(\ba{c}\vec{q} \\ \vec{p}\ea\right)$ 
and $W=\left(\ba{cc}0 &I \\ -I &0 \ea\right)$, a $2d\times 2d$ 
symplectic matrix. This is 
the simplest example where all the operators have a canonically-conjugated 
counterpart. The addition of quadratic polynomials 
\be sp(2d,\Re)=\{\hat{Q}_{jk}\sim  
\hat{q}_j \hat{q}_k,\, \hat{P}_{jk}\sim \hat{p}_j \hat{p}_k,\, 
\hat{R}_{jk}\sim  \hat{q}_j \hat{p}_k\}\label{qua}
\ee
(more precisely, the generators of the symplectic 
group $Sp(2d,\Re)=\{S\in M_{2d\times 2d}/\,\,S^tWS=W\}$) yields the 
Weyl-symplectic group $WSp(2d,\Re)$, whose group law is 
\be\ba{rcl} S''&=&S'S\,, \\ 
\vec{V}''&=&\vec{V}'+S'\vec{V}\,, \\ 
\zeta''&=&\zeta'\zeta e^{\frac{i}{2\hbar} \vec{V}'^t WS'\vec{V}}\,.\ea
\label{ws}
\ee
The unitary irreducible representations of the 
group $WSp(2d,\Re)$ describe a quantum free-like dynamics  
(Galilean particle, harmonic oscillator, etc) when the structure group 
is just $\TT=U(1)$. Other choices of $\TT$ lead to constrained 
(non-linear, in general) quantum dynamics on a submanifold $\Gamma$ of 
the phase space $\Re^{2d}$; for example, the Affine structure subgroup 
$\TT=\tilde{A}(1)$, which is generated by $\{\hat{\vec{q}}^2-r^2\hat{I},\,
\um(\hat{\vec{q}}\hat{\vec{p}}+\hat{\vec{p}}\hat{\vec{q}}),\,\hat{I}\}$,  
leads to a particle moving on the $S^{d-1}$ sphere of radius $r$ (see 
\cite{symplin}). 
Nevertheless,  in comparison with the Heisenberg-Weyl group, 
the Weyl-symplectic group, as such, 
does not incorporate new degrees of 
freedom into the theory unless extra central extensions are considered. 
For example, the appearance of internal degrees of freedom manifests 
itself through a non-trivial transformation 
$\zeta\stackrel{S'}{\rightarrow} \zeta e^{\frac{i}{\hbar}
\xi_{{\small{\rm cob}}}(S',S)}$ of the phase 
under the action of the symplectic group 
$Sp(2n,\Re)\subset WSp(2n,\Re)$ (note 
that this possibility is not realized in the group law (\ref{ws})), 
where $\xi_{{\small{\rm cob}}}(S',S)=\eta(S'S)-\eta(S)-\eta(S')$ is a (pseudo) 
cocycle generated by the linear function 
$\eta(S)=\alpha_{jk}q_{jk}+\beta_{jk}p_{jk}+
\rho_{jk}r_{jk}$ in the coordinates 
$q_{jk},p_{jk},r_{jk}$ of the symplectic group (attached to the Lie-algebra 
generators (\ref{qua})) and $\alpha_{jk}, \beta_{jk},\rho_{jk}$ are 
real parameters. When the cocycle (in fact, coboundary) 
$\xi_{{\small{\rm cob}}}(S',S)$ 
is added to the cocycle $\xi( \vec{V}'S'^t, \vec{V})=
\frac{1}{2} \vec{V}'^t WS'\vec{V}$ in the 
group law (\ref{ws}), i.e. when the phase transform as 
\be\zeta''=\zeta'\zeta e^{\frac{i}{\hbar}\xi( \vec{V}'S'^t, \vec{V})}
e^{\frac{i}{\hbar}\xi_{{\small{\rm cob}}}(S',S)}\,,\ee
it induces the appearance of central charges 
in the Lie-algebra commutators of $sp(2d,\Re)$. 
For example, a non-trivial response  
of the phase under rotations $SU(2)\subset Sp(6,\Re)$, given by the 
generating function  
$\eta(S)=\rho_{12}(r_{12}-r_{21})$, deforms the Lie-algebra of angular 
momentum operators $\hat{L}_i=\epsilon_{ijk}\hat{R}_{jk}$ and leads to the 
appearance of central terms
\be
\left[\hat{L}_1,\hat{L}_2\right]=i\hbar\hat{L}_3+\rho_{12}\hbar\hat{I}
\ee
proportional to the {\it spin} parameter $s=\rho_{12}/\hbar$. 
Thus,  the couple 
$(\hat{L}_1,\hat{L}_2)$ becomes a conjugated 
pair of operators, on an equal footing with $(\hat{q}_j,\hat{p}_j)$; 
in fact, the operators $\hat{L}_j$ are no longer written in terms of the 
basic operators $\hat{q}_j,\hat{p}_k$, since they, themselves, are now  
basic (spin operators) and represent a new degree of freedom of the 
quantum theory given by a unitary irreducible representation of the 
``deformed'' Weyl-symplectic group 
(the ``one'' which incorporates $\xi_{{\small{\rm cob}}}$). 
A similar argument applies for non-trivial responses of the phase under the 
action of the $SL(2,\Re)\subset Sp(2d,\Re)$ subgroup, given by the 
generating function $\eta(S)=\rho \sum_{j=1}^dr_{jj}$, which deforms the 
Lie-algebra of {\it squeezing} operators $\{\hat{Q}=
\um\sum_{j=1}^d\hat{Q}_{jj}, \hat{P}=\um\sum_{j=1}^d\hat{P}_{jj}, 
\hat{R}=\sum_{j=1}^d\hat{R}_{jj}\}$ by introducing a central term
\be
\left[\hat{Q},\hat{P}\right]=i\hbar\hat{R}+\rho\hbar\hat{I}
\ee
proportional to the {\it symplin} (symplectic spin) 
parameter $\bar{s}=\rho/\hbar$ (see \cite{symplin}). The symplin degree 
of freedom shares with spin its internal character, although it possesses  
an infinite number of internal states which correspond to the carrier 
space of the irreducible representations of the non-compact group  
$SL(2,\Re)$ with Bargmann index $k=\bar{s}$. Unlike spin, 
whose physical significance is well 
understood in terms of fermionic and bosonic objects, symplin does 
not seem to fit any known characteristic of the elementary 
particle. Only the (quantum) {\it critical} value of 
$\bar{s}_0=d/2$ (for which an {\it anomalous} 
reduction of the representation is possible \cite{symplin}) seems to have  
a direct physical meaning related to the {\it zero point energy} 
$E_0=\bar{s}_0\hbar\omega$ of a $d$-dimensional  
harmonic oscillator (observable, for example, in the Casimir effect 
\cite{Itzykson}). This 
particular value of $\bar{s}_0=d/2$ tells us that the quantization map 
``\,\,$\hat{{}}\,\,$'' assigns the operator $\hat{R} 
-i\hbar\bar{s}_0\hat{I}=\um\sum_{j=1}^d(\hat{q}_j\hat{p}_j+\hat{p}_j
\hat{q}_j)$ (actually, the {\it symmetrized} operator)  
to the classical function  $r=\sum_{j=1}^dq_jp_j$. This anomalous reduction 
for the critical value of $\bar{s}_0=d/2$, which allows the operator $\hat{R}$ 
to be written in terms of the basic operators $\hat{q}$ and $\hat{p}$, 
is of the same nature 
as the anomalous reduction which allows the Virasoro operators 
$\hat{L}_k$ to be written in terms of the string modes $\hat{\alpha}^\nu_n$, 
more explicitly  
\be
\hat{L}_k=\um g_{\mu\nu} :\sum_n \hat{\alpha}^\mu_{k-n} \hat{\alpha}^\nu_n:
\label{suga}
\ee
(the Sugawara construction \cite{Suga}), for the critical values $c_0=c_0'=d$ 
(dimension of the space-time) of the central extension parameters of the 
Virasoro algebra
\be
\left[\hat{L}_n,\hat{L}_m\right]=\hbar(n-m)\hat{L}_{n+m}+
\frac{\hbar^2}{12}(cn^3-c'n)\delta_{n,-m}\hat{I}\,.\label{vira}
\ee

However, the possibility of a non-critical value of $\bar{s}$ prompts us 
to wonder whether or not quantizing outside the value $\bar{s}_0$ makes 
any sense. The fact is that, the symplin degree of freedom can be 
understood as forming part of a larger set of degrees of freedom 
originating in the free particle and conforming to an extended object 
which proves to generalize other physical systems bearing conformal 
symmetry. More precisely, this 
extended object arises when one is trying to quantize more classical 
observables than those allowed by the well-known {\it no-go} theorems of 
Groenwald and van Hove \cite{GvH,Abraham,Gotay}. Indeed, the standard 
canonical quantization fails to go beyond 
any Poisson algebra containing polynomials in $q$ and $p$ of degree 
greater than two. Ambiguities in quantization arise mainly due to ordering 
problems; for example, the quantization mapping ``\,\,$\hat{{}}\,\,$'' 
is not unique for the classical function $q^2p^2$. From the point of view 
of group quantization, the quantum morphism ``\,\,$\hat{{}}\,\,$'' can be 
distorted because of the appearance of inescapable  terms 
(central ones, in particular) in the quantum Lie-algebra commutators,  
whose classical counterpart 
(Poisson algebra brackets) do not possess. In fact, let us choose the next 
set of classical functions of the harmonic oscillator variables 
$a=\frac{1}{\sqrt{2}}(q+ip),\,a^*=\frac{1}{\sqrt{2}}(q-ip)$ 
(we are using $m=1=\omega$, for simplicity):
\be
L^\alpha_{|n|}=\um a^{2|n|}(aa^*)^{\alpha-|n|+1},\,\,\,
L^\beta_{-|m|}=\um a^{*2|m|}(aa^*)^{\beta-|m|+1},\label{auaral}
\ee
where $n,m,\alpha,\beta\in Z/2$. 
A straightforward calculation from the 
basic Poisson bracket $\{a,a^*\}=i$ provides the following formal 
Poisson algebra:
\be
\{L^\alpha_n,L^\beta_m\}=-i[(1+\beta)n-(1+\alpha)m]L^{\alpha+\beta}_{n+m}\,,
\label{auaralcom}
\ee
which formally generalizes the Poisson algebra $w_\infty$ of functions on a 
two-dimensional phase space \cite{WB} (to `half-integer' indices). 
The (conformal-spin-2) 
generators $L_n\equiv L^0_n$ close the  
Virasoro algebra (\ref{vira}) without central extension, 
and the (conformal-spin-1) generators 
$\alpha_m\equiv L^{-1}_m$ close the non-extended Abelian Kac-Moody 
algebra of ``string-modes''; in general $L^{\alpha}_n$ has 
conformal-spin $N=\alpha+2$ and conformal-dimension $n$ (the eigenvalue of 
$L^0_0$). 

If the analyticity of the classical functions $L^\alpha_n$ is 
taken into account, then one should worry about a restriction 
of the range of the indices $n,\alpha$. The subalgebra of 
polynomial functions $w_\wedge\equiv\{ L^\alpha_n\,,/\,\, 
\alpha-|n|+1\geq 0,\,\,n,\alpha\in Z\}$ closes a 
subalgebra of  (\ref{auaralcom}) and corresponds to the `classical limit' 
$\hbar\rightarrow 0$ of the so-called 
{\it wedge} subalgebra $W_\wedge$ \cite{Pope} (isomorphic to 
$sl(\infty,\Re)$ algebra), which is a particular case 
($\tau(0)$)  of $SL(2,\Re)$ tensor-operator algebras denoted 
collectively by $\tau(\mu)$ [the quotient of the enveloping algebra 
${\cal U}(sl(2,\Re))$ by the ideal generated by $Q-\mu$, where 
$Q$ is the Casimir of $sl(2,\Re)$ and $\mu$ is related to the symplin 
$\bar{s}$ by $\mu=\frac{\bar{s}}{2}(\frac{\bar{s}}{2}-1)$]. The tensor 
algebras $\tau(\mu)$ can be considered as ``quantum'' deformations 
(mainly due to ordering ambiguities) 
\bea
\l[\hat{L}^\alpha_n,\hat{L}^\beta_m\r]&=&
\sum_{j=0}{\hbar^{2j+1}f^{\alpha\beta}_{2j}(n,m;\mu)
\hat{L}^{\alpha+\beta-2j}_{n+m}}\,,\label{deform}\\
f^{\alpha\beta}_{0}(n,m;\mu)&=&[(1+\beta)n-(1+\alpha)m]\,,\nn
\eea
of $w_\wedge$, and their essence  
can be captured in a classical construction by extending the Poisson 
bracket $\{\cdot,\cdot\}$ to the Moyal bracket $\{\cdot,\cdot\}_{M}$ 
\cite{Moyal} of functions on the co-adjoint orbits of $SL(2,\Re)$ (the 
wedge algebra $W_\wedge$ corresponds to the case where the co-adjoint 
orbit is a cone $\Re^+\times S^1$). For the critical value of 
the symplin $\bar{s}_0=\um$, the algebra 
$\tau(\mu=-\frac{3}{16})$ is the ``symplecton'' algebra of Biedenharn 
and Louck (see \cite{Pope} and references therein), and the limit 
$\mu\rightarrow\infty$ corresponds to the area-preserving diffeomorphisms 
$SDiff(H)$ of a two-dimensional hyperboloid $H$. 

Quantum deformations (\ref{deform}) of $w_\wedge$ do not introduce true 
central extensions (the situation changes when half-integer values of the 
indices are allowed). The inclusion of central terms in (\ref{deform}) 
requires the formal extension $W_\infty(\mu)$ of $\tau(\mu)$ beyond the wedge 
$|n|\leq \alpha+1$ \cite{Pope}, that is, 
the consideration of non-polynomial functions (\ref{auaral}). If  
we use polar coordinates $a=re^{i\theta}$, then the functions (\ref{auaral}) 
acquire the form $L^{\alpha}_n=r^{2(1+\alpha)}e^{2in\theta}$, which are 
single-valued for all values of $\alpha,n\in Z/2$ with the condition 
$\alpha\geq -1$ (strictly positive conformal-spin). Central extensions of 
(\ref{deform}) ---outside the wedge--- of the form 
\be 
\Sigma(\hat{L}^\alpha_n,\hat{L}^\beta_m)=\hbar^{2\alpha+2}c_\alpha(n;\mu)
\delta^{\alpha,\beta}\delta_{n+m,0}\hat{I}\label{cext}
\ee
are known for the particular cases:    
$W_{\infty}\equiv W_{\infty}(0)$ and $W_{1+\infty}\equiv 
W_{\infty}(-\frac{1}{4})$. It is also precisely 
for these specific values of $\mu=0,-\frac{1}{4}$ that  
the sequence of terms 
on the right-hand side of (\ref{deform}) turns out to be zero whenever 
$\alpha+\beta-2j\leq 0$ and $\alpha+\beta-2j\leq -1$, respectively, and 
therefore  $W_{\infty}$ (resp. $W_{1+\infty}$) 
can be consistently truncated to a 
closed algebra containing only those generators with conformal-spins $\geq 2$ 
(resp. $\geq 1$). The aforementioned central term provides extensions 
for all (positive) conformal-spin currents, in particular the standard 
central extension (\ref{vira}) for the Virasoro sector, and the central 
extension $[\hat{\alpha}_n,\hat{\alpha}_m]=nc\delta_{n+m,0}\hat{I}$ for the 
Abelian Kac-Moody subalgebra of conformal-spin-1 `string modes', when 
$\mu=-\frac{1}{4}$ (see \cite{Pope}).

The algebras $w_{\infty}$ and $w_{1+\infty}$ generalize the underlying 
Virasoro gauged symmetry 
of the light-cone two-dimensional induced gravity discovered by Polyakov 
\cite{Poly}, and induced actions for these $w$-gravity 
theories have been proposed in 
\cite{Pope1}. For them, the quantization procedure 
deforms $w$ to $W$ symmetry due to the presence of anomalies. Also, hidden 
$SL(\infty,\Re)$ and $GL(\infty,\Re)$ Kac-Moody symmetries exist for 
$W_{\infty}$ and $W_{1+\infty}$ gravity (see \cite{Pope2} and \cite{Shen} 
for a review), respectively, 
generalizing the hidden $SL(2,\Re)$ Kac-Moody symmetry of Polyakov's 
induced gravity \cite{Poly2}. 

Thus, $W_\infty(\mu)$ theories for the `critical' values $\mu=0$ and 
$\mu=-\frac{1}{4}$ are fairly well understood (also induced $W_\infty$ 
gravity has being obtained as a WZNW model \cite{Nissimov}). 
In contrast, 
gauge theories of $W_\infty(\mu)$-algebras, for general $\mu$, are said to 
be less interesting because they include, in general, {\it negative} 
conformal-spins $N=\alpha+2\leq 0$. However, the consideration of 
negative conformal-spins provide new central extensions 
which, as far as we know, have not been considered in the literature. 
In fact, in addition to the  Virasoro-sector central term 
\be
\Sigma_0(\hat{L}^\alpha_n,\hat{L}^\beta_m)=\hbar^2 
\frac{c}{12}(n^3-n)\delta^{\alpha,0}\delta^{\beta,0}\delta_{n+m,0}\hat{I}\,, 
\ee
which is the only possible central extension (\ref{cext}) 
of (\ref{auaralcom}) as such  
for positive conformal-spins $N>0$ (see \cite{Pope}), extra central 
extensions providing new couples of conjugated generators
\be
\left[\hat{L}^\alpha_n, \hat{L}^{-1-\alpha}_{-n}\right]=
\hbar n\hat{L}^{-1}_0+O(\hbar^2)=\um \hbar n\hat{I} +O(\hbar^2)
\ee
are possible when negative conformal-spins are considered. 
Indeed, the generator $\hat{L}^{-1}_0$ (the quantum counterpart of the 
the classical constant function ${L}^{-1}_0=\um$) commutes with all 
the other generators and plays the role of the central generator $\hat{I}$. 
Classical functions $L^\alpha_n$ with negative conformal-spins $\alpha<-1$ 
are singular at the origin $aa^*=0$. 
Now, from a quantum point of view, each 
operator $\hat{L}^\alpha_n$ with $\alpha<-1$ 
has a conjugated counterpart 
$\hat{L}^{-1-\alpha}_{-n}$ (related to a non-singular classical 
function) and no longer needs  to be written in 
terms of  other effectively basic operators such as 
$\hat{L}^{-\um}_{\um}=\um \hat{a}$ and $\hat{L}^{-\um}_{-\um}=\um 
\hat{a}^{\dag}$, thus avoiding problems of lack of analyticity. 
Even more, the operator $\hat{L}^0_0$ 
associated with the classical function 
$L^0_0=\um aa^*$ is never zero because of the existence of a 
``zero point energy'' (critical value of the symplin degree of freedom) 
associated with the inherent ordering ambiguity;  
also, square roots of $aa^*$ are a minor 
harm from the quantum point of view. 

In other words, the removal of the origin 
from the original symplectic manifold 
(leading to a punctured complex plane) 
creates new cohomology on the Poisson algebra of functions, 
i.e. new couples of conjugated generators (in fact, an infinite amount) 
emerge when the topology of the 
phase-space turns less trivial ($\Re^2\rightarrow \Re^+\times S^1$) and, as 
a result,  new quantum {\it extended} objects possessing  
conformal symmetry appear. 
The field content of these  extended objects 
and the possible deformations (renormalizations) caused by the quantization 
procedure are being investigated from a GAQ framework \cite{progress1}. An 
explicit expression for the corresponding invariant action functional 
$S$ can be given from the integral $S=\int{\Theta}$ of the one-form $\Theta$ 
in (\ref{thetagen}), which generalizes the 
Poincar\'e-Cartan form $\Theta_{PC}$ of Classical Mechanics for the 
quantum phase term $-i\hbar\zeta^{-1}d\zeta$, along the trajectories of 
the l.i.v.f. in the characteristic subalgebra (\ref{cara}). 
This way of constructing field models 
from two-cocycles (see Ref. \cite{coadjoint}) differs from the analysis in 
related works based on the  coadjoint-orbit approach 
\cite{GvH,Abraham,coadjoint1,coadjoint2}.  

Also,  the study of tensor-operator algebras of more general 
simple groups, their precise 
relation with  the Poisson algebra of functions on the corresponding 
co-adjoint orbits and the physical meaning of its extensions `beyond 
the wedge' is in progress \cite{progress2}. For the case of $SU(p)$, 
the generators $L^\alpha_n$ are now labelled by a $(p-1)$-dimensional ---the 
rank of $SU(p)$--- vector $\alpha=(\alpha_1,\dots,\alpha_{p-1})$, which is 
taken to lie on an integral lattice, and an upper-triangular 
$p\times p$ matrix $n$, with $p(p-1)/2$ integral entries 
---the dimension of the biggest coadjoint orbit--- 
(see \cite{progress2} for more details).

 Let us see how  $W_{1+\infty}$ symmetry also arises in a two-dimensional 
electron gas in a perpendicular magnetic field, and its relevance in the 
classification of all the universality classes of {\it incompressible 
quantum fluids} and the identification of the quantum numbers of the 
excitations in the Quantum Hall Effect (QHE).

\section{Quantum Hall Universality Classes and $W_{1+\infty}$ 
Symmetry}\label{Hall}

QHE can be considered the paradigm of the two-dimensional quantum physical 
systems. The extreme precision of the rational values  of the Hall 
conductivity $\sigma_{x_1x_2}$ (in $e^2/h$ units) suggests that the dynamics 
of planar electrons  in a perpendicular magnetic field 
$H$ is constrained by symmetry and topology. 
In fact, {\it magnetic translations} 
$x_j=-\epsilon_{jk}q_k+\frac{p_j}{m\omega_c},\, j=1,2$ 
($\omega_c=\frac{eH}{mc}$ denotes 
the cyclotron frequency), which are 
standard space translations combined with gauge transformations, are the 
most important underlying symmetry of this system. The commutator 
of two infinitesimal magnetic translations is 
$\left[\hat{x_j},\hat{x_k}\right]=i\frac{m\omega_c}{\hbar}\epsilon_{jk}
\hat{I}$, which reproduces the Heisemberg-Weyl commutation relations  
given in (\ref{HW}). This symmetry is responsible for the infinite 
degeneracy of the Landau levels, degeneracy 
that becomes {\it finite} when boundary 
conditions (for example, periodicity conditions) are imposed. This 
constrained symmetry can be described by means of the Heisemberg-Weyl 
group $\TG$ (\ref{HWlaw}) in $d=1$, that is:
\be
x_j''=x_j'+x_j,\,\,\,\,
\zeta''=\zeta'\zeta e^{i\frac{m\omega_c}{2\hbar}(x_1'x_2-x_1x_2')}\,,
\ee
where finite translations $x_j\rightarrow x_j+k_jl_j$ 
by an integer amount of $l_j$ (spatial period) play the role of the 
structure subgroup $\TT\sim Z\times Z\times U(1)\subset \TG$ 
(see \cite{FracHall} for more details).  
One can easily check from 
the last group law that, for integer values of  
$\phi=\frac{m\omega_cl_1l_2}{2\pi\hbar}=n$, 
the structure subgroup $\TT$ is Abelian 
(i.e. $\TT$ is a trivial central extension $\TT=Z\times Z\times U(1)$), 
so that all the constraints are first-class,  
and the unitary irreducible representations of $\TG$ are 
$n$-dimensional (in contrast with the infinite-dimensional character 
of the $\TT=U(1)$ case). The condition $\phi=n$ represents a quantization 
of the magnetic flux through the torus surface (the quotient  
$\Re\times\Re/Z\times Z$)  in the 
same manner as in the Dirac monopole case. Nevertheless, this 
quantization condition is not strictly necessary and fractional 
values of the flux $\phi=\frac{n}{r}$ are also allowed 
(the irrational case is more involved and 
requires techniques from Non-Commutative Geometry \cite{Connes}). 
The structure subgroup 
$\TT$ is non-Abelian for this case (it is a non-trivial central extension), 
so that constraints are second-class, and the unitary irreducible 
representations of $\TG$ are $n\times r$-dimensional (``spinorial''-like 
representations) \cite{FracHall}. 

There is a reciprocal relation between the 
flux $\phi$ and the {\it filling factor}  $\nu=h\sigma_{x_1x_2}/e^2$, which  
is a very stable (topological) number that 
appears with {\it odd} denominators for fermionic carrying systems (perhaps 
modular invariance could explain this `odd denominator rule' \cite{modular}). 
Very accurate trial wave functions were proposed by Laughlin 
(see \cite{Prange} for a review) to describe the behavior of the ground state 
for different values of $\nu$. 
These are macroscopic quantum states with uniform density and a gap for 
density fluctuations, which lead to the key idea of {\it two-dimensional 
incompressible quantum fluids}, and can be classically thought of as 
{\it droplets of liquid} without density waves. Thus, at the classical level, 
all possible configurations of droplets of incompressible fluid 
({\it edge excitations}) can be generated by area-preserving diffeomorphisms 
from a reference droplet, that is, by applying generators of $w_{1+\infty}$ 
on the ground-state distribution function $\Omega$. This dynamical 
symmetry, which has been identified by \cite{Capelli}, traces back to the 
extra canonical transformations of the four-dimensional phase space that 
leave invariant the Hamiltonian; that is, functions like (\ref{auaral}) of 
magnetic translations $a=\um(x_2+ix_1)$. 

The quantization of these edge excitations can be achieved by studying 
the irreducible, unitary highest-weight representations of 
$W_{1+\infty}$ (a complete classification has 
been given in \cite{Kac}), which correspond to $(1+1)$-dimensional 
effective conformal field theories. In this formalism, the incompressible 
quantum fluid ground state $|\Omega\rangle$  appears as a highest-weight 
state satisfying $\hat{L}^\alpha_n|\Omega\rangle=0,\,\,\forall n\geq 0,\,\,
\alpha\geq -1$. Particle-hole edge excitations above the ground state 
are obtained by applying generators with negative mode index, 
$\hat{L}^\alpha_{-|n|}$, to $|\Omega\rangle$. These highest-weight 
conditions are automatically fulfilled from the GAQ point of view. Indeed, 
according to general settings \cite{Ramirez,conforme}, the vacuum of a 
quantum theory defined through a quantizing group $\TG$  
must be annihilated by the right-version of the polarization subalgebra 
dual to ${\cal G}_p$ in (\ref{pola}). For the case of the 
quantizing algebra $\tilde{{\cal G}}=W_{1+\infty}$, the characteristic 
subalgebra (\ref{cara}) is  
 ${\cal G}_c=\{\hat{X}^\alpha_0,\,\,\alpha\geq -1\}$ ($\hat{X}$ means the 
left-invariant vector field counterpart of $\hat{L}$), which can be enlarged 
to a polarization subalgebra ${\cal G}_p=\{\hat{X}^\alpha_n,
\,\,\alpha\geq -1,\,n<0\}$. The polarization conditions (\ref{pola}) 
$\hat{X}\psi=0,\,\,\hat{X}\in {\cal G}_p$ reduce the representation 
(\ref{repre}) of $\TG$ on $U(1)$-equivariant 
wave functions (\ref{tequiv}).   

The point is that one can 
presumably characterize {\it any} quantum incompressible fluid as a  
$W_{1+\infty}$ theory, so that the hierarchy problem 
(that is, the classification of stable ground states and their 
excitations corresponding to all observed {\it plateaus} 
or filling fractions) reduces to a 
complete classification of $W_{1+\infty}$ theories (see \cite{Capelli} 
for more details). 

The general idea that central extensions of the symmetry group of a 
physical system provide new (quantum) degrees of freedom is also 
applicable to the appearance of mass (in a {\it non}-standard way) 
in Yang-Mills quantum theories, 
as we are going to see now.

\section{Group Quantization of Yang-Mills Theories: a Cohomological 
Origin of Mass}

Some of the essential issues  we have discussed in Sec. \ref{extended} about 
the irreducible representations of $WSp(2d,\Re)$ bearing internal 
degrees of freedom, will be useful in showing how mass can enter Yang-Mills 
theories through central (pseudo) extensions of the 
corresponding gauge group. This 
mechanism does not involve extra (Higgs) scalar particles and could 
provide new clues for the better understanding of the nature of the 
Symmetry Breaking Mechanism. 
We are going to outline the essential points and refer the 
interested reader to Refs. \cite{ym,gtp} for further information. 

Let us denote by $A^\mu(x)=r^a_bA^\mu_a(x)T^b,\,\mu=0,...,3; a,b=1,...,n$ the 
Lie-algebra valued vector potential attached to a non-Abelian gauge 
group which, 
for simplicity, we suppose to be unitary, say $T={\rm Map}(\Re^4,SU(N))=
\{U(x)=\exp{\varphi_a(x)T^a}\}$, where $T_a$ are the generators of $SU(N)$,  
which satisfy the commutation relations $[T_a,T_b]=C_{ab}^cT_c$, and 
the coupling constant matrix $r^a_b$ reduces to a multiple of the identity 
$r^a_b=r\delta^a_b$. 
We shall also make partial use of the gauge freedom to set the 
temporal component $A^0=0$, so that the Lie-algebra valued 
electric field is simply $E^j(x)\equiv r^a_bE^j_a(x)T^b=-\dot{A}^j(x)$. 
In this case, there is still a residual gauge 
invariance $T={\rm Map}(\Re^3,SU(N))$ (see \cite{Jackiw}).

The proposed (infinite dimensional) quantizing group for quantum Yang-Mills 
theories will be a central extension $\TG$ of 
$G=(G_A\times G_E)\times_s T$ (semi-direct product of the cotangent 
group of the Abelian group of Lie-algebra valued vector potentials 
and the non-Abelian 
gauge group $T$) by $U(1)$. More precisely, the group law for $\TG$, $\tg''=\tg*\tg$, 
with $\tg=(A^j_a(x),E^j_a(y),U(z);\zeta)$, 
 can be explicitly written as (in natural 
units $\hbar=1=c$):
\bea
U''(x)&=&U'(x)U(x)\,,\nn\\
\vec{A}''(x)&=&\vec{A}'(x)+U'(x)\vec{A}(x)U'(x)^{-1}\,,\nn\\
\vec{E}''(x)&=&\vec{E}'(x)+U'(x)\vec{E}(x)U'(x)^{-1}\,,\nn\\
\zeta''&=&\zeta'\zeta\exp\left\{-\frac{i}{r^2}\sum_{j=1}^2
\xi_j(\vec{A}',\vec{E}',U'|\vec{A},\vec{E},U)\right\}\,;
\label{ley}\\
\xi_1(g'|g)&\equiv& \int{{d}^3x\,{\rm tr}\left[\,\left(\ba{cc} \vec{A}' & 
\vec{E}'\ea\right) W \left(\ba{c} U'\vec{A}U'^{-1} \\ 
U'\vec{E}U'^{-1} \ea\right)\right]}\,,\nn\\
\xi_2(g'|g)&\equiv& \int{{d}^3x\,{\rm tr}\left[\,\left(\ba{cc} 
\nabla U'U'^{-1} & 
\vec{E}'\ea\right) W \left(\ba{c} U'\nabla UU^{-1}U'^{-1} \\ 
U'\vec{E}U'^{-1} \ea\right)\,\right]}\,,\nn 
\eea
\ni where $W=\left(\ba{cc} 0 & 1 \\ -1 & 0\ea\right)$ is a symplectic 
matrix and we have split up the cocycle $\xi$ into two distinguishable 
and typical cocycles $\xi_j,\,\,j=1,2$. The first cocycle  
$\xi_1$ is meant to provide {\it dynamics} 
for the vector potential, so that the couple $(A,E)$ 
corresponds to a canonically-conjugate pair of coordinates.  
The second cocycle $\xi_2$, 
the {\it mixed} cocycle, provides a non-trivial (non-diagonal) action 
of the structure subgroup $\TT$ on vector potentials and determines 
the number of degrees of freedom of the constrained theory; in fact, it 
represents the ``quantum'' counterpart  
of the ``classical'' unhomogeneous term $U(x)\nabla U(x)^{-1}$ we miss 
at the right-hand side of the gauge transformation of $\vec{A}$ 
(second line of 
(\ref{ley})), that is, the vector potential $\vec{A}$ 
has to transform homogeneously 
under the action of the gauge group $T$ in order to define a proper 
group law, whereas the inhomogeneous term $U(x)\nabla U(x)^{-1}$ modifies  
the {\it phase} $\zeta$ of the wave function according to $\xi_2$ 
(see \cite{gtp,ym} for a covariant form of this ``quantum'' 
transformation). 

To make more explicit the intrinsic 
significance of these two quantities $\xi_j\,,\,\, j=1,2$, let us 
calculate the non-trivial Lie-algebra commutators of the right-invariant 
vector fields (that is, the generators of the left-action 
$L_{\tg'}(\tg)=\tg'*\tg$ of $\TG$ on itself) from  the group law (\ref{ley}). 
They are explicitly: 
\bea
\l[\hat{A}^j_a(x), \hat{E}^k_b(y)\r]&=&
i\delta_{ab}\delta^{jk}\delta(x-y)\hat{I}\,,\nn\\
\l[\hat{A}^j_a(x), \hat{\varphi}_b(y)\r]&=&-iC_{ab}^c\delta(x-y) \hat{A}^j_c(x)
-\frac{i}{r}\delta_{ab}\partial^j_x\delta(x-y)\hat{I}\,,\label{YM}\\
\l[\hat{E}^j_a(x), \hat{\varphi}_b(y)\r]&=&-iC_{ab}^c\delta(x-y) 
\hat{E}^j_c(x)\nn\\
\l[\hat{\varphi}_a(x), \hat{\varphi}_b(y)\r]&=&-iC_{ab}^c\delta(x-y) 
\hat{\varphi}_c(x) \,,\nn
\eea
which agree with those of Ref. \cite{Jackiw}. 

The unitary irreducible 
representations of $\TG$ with structure subgroup $\TT=T\times U(1)$ (a direct 
product for this case) represent a quantum theory of 
$n=N^2-1={\rm dim}(SU(N))$ 
interacting massless vector bosons. Indeed, we start with $f=3n$ field degrees 
of freedom, corresponding to the basic operators 
$\{\hat{A}^j_a(x),\hat{E}^j_a(x)\}$ 
(the ones that have a conjugated 
counterpart); the constraints (\ref{tequiv}) 
provide $c=n$ independent restrictions 
$\hat{\varphi}_a(x)\psi=0,\,\, a=1,\dots, n$ (the quantum implementation 
of the non-Abelian Gauss law), since 
they are first-class constraints and we choose the trivial representation 
$D_{\TT}^{(\epsilon)}(\tg_t)=1,\,\,\forall \tg_t=(0,0,U(x);1)\in T$, 
restrictions  which lead to $f_c=f-c=2n$ field 
degrees of freedom corresponding to an interacting theory of $n$ massless 
vector bosons. 

However, more general representations 
$D_{\TT}^{(\epsilon)}(U)=e^{i\epsilon_{{}_U}}$ 
can be considered when we impose 
additional boundary conditions like $U(x)
\stackrel{x\rightarrow\infty}{\longrightarrow}\pm I$, that is, when we 
compactify the space $\Re^3\rightarrow S^3$ so that the gauge group $T$ 
falls into disjoint homotopy classes $\{U_l\,,\,
\epsilon_{{}_{U_l}}=l\vartheta\}$ 
labeled by integers $l\in Z=\pi_3(SU(N))$ (the third homotopy group). 
The index $\vartheta$ (the {\it $\vartheta$-angle} 
\cite{Jackiwtheta}) parametrizes 
{\it non-equivalent quantizations}, as the Bloch momentum $\epsilon$ does  
for particles in periodic potentials, where the wave function acquires 
a phase $\psi(q+2\pi)=e^{i\epsilon}\psi(q)$ after a translation of, 
let us say, $2\pi$. The 
phenomenon of non-equivalent quantizations can also be reproduced by 
keeping the constraint condition  $D_{\TT}^{(\epsilon)}(U)=1$ unchanged 
at the price of introducing a new (pseudo) cocycle 
$\xi_\vartheta$ which is added to the 
previous cocycle $\xi=\xi_1+\xi_2$ in (\ref{ley}). The generating function 
$\eta_\vartheta$ of $\xi_\vartheta$ is 
\be
\eta_\vartheta(g)=\vartheta\int{d^3x\, {\cal C}^0(x)}\,,\;\;\;\;
 {\cal C}^\mu=-\frac{1}{16\pi^2}\epsilon^{\mu\alpha\beta\gamma}{\rm tr}
({\cal F}_{\alpha\beta}{\cal A}_\gamma-\frac{2}{3}{\cal A}_\alpha 
{\cal A}_\beta {\cal A}_\gamma)\,,
\ee
where ${\cal A}\equiv A+\nabla UU^{-1}$ and ${\cal C}^0$ is 
the temporal component of the {\it Chern-Simons secondary characteristic 
class} ${\cal C}^\mu$, which is the vector whose divergence equals the 
Pontryagin density  ${\cal P} = \partial_\mu {\cal C}^\mu = -\frac{1}{16\pi^2} 
{\rm tr}  ({}^*{\cal F}^{\mu\nu} {\cal F}_{\mu\nu})$ 
(see \cite{Jackiw}, for instance). 
Like some total derivatives (namely, the Pontryagin density), 
which do not modify 
the classical equations of motion when added to the Lagrangian but have a 
non-trivial effect in the quantum theory, pseudo-cocycles like 
$\xi_\vartheta$ give rise to non-equivalent quantizations when 
the topology of the space is affected by the imposition of 
certain boundary conditions (``compactification of the space''), 
even though they are trivial cocycles of  the ``unconstrained'' theory. 
The phenomenon of non-equivalent quantizations can be also sometimes 
understood as a {\it Aharonov-Bohm-like effect} (an effect experienced by the 
quantum particle but not by the classical particle) and 
$d\eta(g)=\frac{\partial\eta(g)}{\partial g^j}d g^j$ can be seen 
as an {\it induced gauge connection} (see \cite{FracHall} for the example 
of a super-conducting ring threaded by a magnetic flux) which modifies  
momenta according to the minimal coupling.

We can also go further and consider more general representations 
$D_{\TT}^{(\epsilon)}$ of $\TT$ (in particular, non-Abelian representations) 
by adding extra pseudo-cocycles to $\xi$. This is the case of 
\be
\xi_\lambda(g'|g)\equiv -2\int{{d}^3x\, {\rm tr}[ 
\lambda\left(\log (U'U)-\log U'-\log U\right)]}\,,\nn
\ee
which is generated by $\eta_\lambda(g)=-2\int{{d}^3x\, {\rm tr}[
\lambda\log U]}$, where $\lambda=\lambda^aT_a$ is a matrix carrying some 
parameters $\lambda^a$ which actually characterize the representation of 
$\TG$. In fact, this pseudo-cocycle alters the gauge group commutators 
and leads to the appearance of new central terms at the right-hand side 
of the last equation in (\ref{YM}), more explicitly:
\be
\l[\hat{\varphi}_a(x), \hat{\varphi}_b(y)\r]=-iC_{ab}^c\delta(x-y) 
\hat{\varphi}_c(x) 
-iC_{ab}^c\frac{\lambda_c}{r^2}\delta(x-y)\hat{I}\,.\label{masin}
\ee
For this case, new couples of generators 
become conjugated  
because of the appearance of these new 
central terms proportional to the  parameter $\lambda_c$ at the 
right-hand side of the commutators  of 
the gauge generators $\hat{\varphi}_a(x)$. That is, new basic operators 
$(\hat{\varphi}_a(x), \hat{\varphi}_b(x))$, such that 
$C_{ab}^c\lambda_c\not=0$, 
join the previous $(\hat{A}^j_a(x),\hat{E}^j_a(x))$, since they 
do not need to be written in terms of them. Constraints are second-class 
for this case and only the set 
${\cal T}_c=\{\hat{\varphi}_a(x),\,/\,\,C_{ab}^c\lambda_c\not=0\,\,
\forall b\}$ (that is, 
the subalgebra of non-dynamical operators of $\TT$),  together with 
half of the dynamical operators of $\TT$ (that is, a polarization 
subgroup $T_p$ of $\TT$),  can be consistently imposed 
as constraints. These facts lead to an increasing of field degrees of freedom 
for the constrained theory, with regard to the former (massless) case. 
In fact, the irreducible representations of the algebra (\ref{YM}) with the 
new commutator (\ref{masin}) define an interacting theory of 
$n_c={\rm dim}(T_c)$ massless vector bosons, 
corresponding to the unbroken gauge subgroup $T_c\subset\TT$ with 
Lie algebra ${\cal T}_c$,  and $n-n_c$ massive-like  
vector bosons with mass cubed $m_c^3=\lambda_c$ 
(see \cite{ym,gtp} for more details). 
 
Pseudo-cocycle parameters such as $\lambda_c$ are usually hidden 
in a redefinition of the generators 
involved in the pseudo-extension 
$\hat{\varphi}_c(x)+\frac{\lambda_c}{r^2}\equiv \hat{\varphi}_c'(x)$, as it 
happens for example with the parameter $c'$ in the Virasoro algebra 
(\ref{vira}), which is a redefinition of $\hat{L}_0$. However, whereas the 
vacuum expectation value $\langle 0_\lambda|
\hat{\varphi}_c(x)|0_\lambda\rangle$ is 
zero,\footnote{it can be easily proven taking into account that the vacuum is 
annihilated by the right version of the polarization subalgebra 
dual to ${\cal G}_p$ \cite{conforme}; also, 
$\hat{\varphi}_c=\xr{\varphi_c}$ is 
always in ${\cal T}_p$; that is, it is zero on constrained wave functionals 
$\Psi_{{\small{\rm phys.}}}$, including the physical vacuum.} the vacuum 
expectation value $\langle 0_\lambda|
\hat{\varphi}_c'(x)|0_\lambda\rangle=\lambda_c/r^2$ 
of the redefined operators $\hat{\varphi}_c'(x)$ is non-null and proportional 
to the cubed mass in the `direction' $c$ of the unbroken gauge 
symmetry $T_c$, which depends on the particular choice of 
the mass matrix $\lambda$. Thus, the effect of the pseudo-extension 
manifests also in a different choice of a vacuum in which some 
gauge operators have a non-zero expectation value. 
This fact reminds us of the Higgs mechanism 
in non-Abelian gauge theories, where the Higgs fields point to the direction 
of the non-null vacuum expectation values. However, the spirit of the  
Higgs mechanism, as an approach to supply mass, and the one 
discussed here are 
radically different, even though they have some common characteristics. 
In fact, we are not making use of extra scalar fields in the theory to 
provide mass to the vector bosons, but it is the gauge group itself 
that acquires dynamics for the massive case and transfers degrees of freedom 
to the vector potentials to form massive vector bosons. 
Thus, the appearance of 
mass seems to have a {\it cohomological origin}, beyond any introduction 
of extra scalar particles (Higgs bosons). The physical implications 
of this alternative approach deserve further study, although some 
important steps have been already done (see \cite{empro,ym,gtp}). 
Also, it would be worth exploring the richness of the case 
$SU(\infty)$ (infinite number of colours), the Lie-algebra of which 
is related to the (infinite-dimensional) Lie-algebra of area preserving 
diffeomorphisms of the torus $SDiff(T^2)$, for which 
{\it true} central extensions 
\be
\l[T_{\vec{m}},T_{\vec{n}}\r]=(\vec{m}\times\vec{n})T_{\vec{m}+\vec{n}}+
\vec{\lambda}\cdot\vec{m}\delta_{\vec{m}+\vec{n},0}\hat{I},\;\;\;\;
\vec{m},\vec{n}\in Z\times Z
\ee
exist (see e.g. \cite{Fairlie}; see also \cite{Kogan} for a change basis 
relating the generators $T_{\vec{m}}$ and $L^\alpha_n$ in Sec. 
\ref{extended},\ref{Hall}). 
Like in String Theory, 
the appearance of central terms in the constraint subalgebra 
does not spoil gauge invariance but 
forces us to impose a polarization subgroup $T_p$ of $\TT$ only 
(namely, the `positive modes' $\hat{L}_{n\geq 0}$ of the Virasoro algebra) 
as restrictions (\ref{tequiv}) on physical wave functions; for this case, 
constraints are said to be of second-class.

\section{Vacuum Radiation and Symmetry Breaking in Conformally Invariant 
Quantum Field Theory}\label{conformesec}

The quantum theory demands an extra requirement of a symmetry in 
comparison with the classical theory, namely the requirement of 
{\it unitarity}, and this proves to be the reason for some breakdown in the 
quantum implementation of some classical symmetries of certain  physical 
systems. For example, the conformal group $SO(4,2)$ has 
always been recognized as a symmetry of the 
Maxwell equations for {\it classical} electro-dynamics 
\cite{Maxwell}, and more recently considered as 
an invariance of general, non-Abelian, massless gauge field theories at 
the classical level; however, the {\it quantum} theory raises, in general, 
serious problems in 
the implementation of conformal symmetry 
 and much work has been devoted to 
the study of the physical reasons for that (see e.g. Ref. \cite{Fronsdal}). 
Basically, the main trouble 
associated with this quantum symmetry (at the second quantization level) 
lies in the difficulty of finding 
a vacuum for massless fields,  which is {\it stable} under special conformal 
transformations acting on 
the Minkowski space in the form:
\be
x^\mu \rightarrow {x'}^\mu=\f{x^\mu+c^\mu
x^2}{\sigma(x,c)},\;\;\;\sigma(x,c)=1+2c x+c^2 x^2
\label{confact}.
\ee
\ni These transformations, which can be interpreted as transitions to systems 
of relativistic, uniformly accelerated observers (see e.g. Ref. \cite{Hill}), 
cause {\it vacuum radiation}, a phenomenon 
analogous to the Fulling-Unruh effect
\cite{Fulling,Unruh} in a non-inertial reference  frame. To be more precise, 
if $a(k),a^+(k)$ are the Fourier components of a scalar massless field 
$\phi(x)$, satisfying the equation 
\be 
\eta^{\mu\nu}\p_{\mu}\p_\nu \phi(x)=0\,,
\ee
\ni  then the Fourier components 
$a'(k), {a'}^+(k)$ of the transformed field 
$\phi'(x')=\sigma^{-l}(x,c)\phi(x)$ by (\ref{confact}) ($l$ being the
conformal dimension) are expressed in terms of both $a(k),a^+(k)$ through a 
Bogolyubov transformation
\be
a'(\lambda)=\int{dk\l[ A_\lambda(k)a(k)+B_\lambda(k)a^+(k)\r]}\,. \label{bog}
\ee
\ni In the second quantized theory,  the vacuum states defined by the
conditions $\hat{a}(k)|0\rangle   =0$ and $\hat{a}'(\lambda)|0'\rangle   =0$, 
are not identical if the coefficients $B_\lambda(k)$ in (\ref{bog})
are not zero. In this case the new vacuum has a non-trivial content of 
untransformed particle states.
 
This situation is always present when quantizing field theories in
curved space as well as in flat space, whenever some kind of global mutilation
of the space is involved. This is the case of the 
natural quantization in Rindler
coordinates \cite{Fulling}, which leads to a quantization inequivalent to the
normal Minkowski quantization, or that of a quantum field in a box, where a
dilatation produces a rearrangement of the vacuum \cite{Fulling}. 
Nevertheless,
it must be stressed that the situation for SCT is more peculiar. 
The rearrangement of the vacuum in a massless
QFT due to SCT, even though they are a 
{\it symmetry} of the classical system, behaves as if the conformal group were 
{\it spontaneously broken}, and this fact can be 
interpreted as some sort dynamical 
{\it anomaly}.

Thinking of the underlying reasons for this anomaly, we are tempted to make
the singular action of the transformations (\ref{confact}) in Minkowski space
responsible for it, as has been in fact pointed out in \cite{rusos}. 
However, a
deeper analysis of the interconnection between symmetry and quantization  
reveals a more profound obstruction to the 
possibility of {\it unitarily}  implementing  
STC  in a  generalized Minkowski space (homogeneous space of the 
conformal group), free 
from singularities. 
This obstruction is traced back to the 
impossibility of representing the entire 
$SO(4,2)$ group {\it unitarily} and irreducibly on a space 
of functions depending 
arbitrarily on $\vec{x}$ (see e.g.  Ref. \cite{Fronsdal}), so that a Cauchy 
surface determines the evolution in 
time. Natural representations, however, can be constructed by means of wave 
functions having support on the whole space-time and evolving in some kind of 
{\it proper time} (related to a dilatation coordinate $\delta$). 
It has been proved (see \cite{conforme}) that 
{\it unitary} irreducible representations of the conformal 
group require the generator $\hat{P}_0$ of time translations to have dynamical 
character (i.e., it has a conjugated pair), as happens with 
the spatial component $\hat{P}_1$, due to the appearance 
of a central term $\hat{I}$ in the quantum commutators  
\be
[\hat{P}_\mu,\hat{K}_\mu]= 
-\eta_{\mu\mu}(2\hat{D}+4N\hat{I})
\ee
\ni  ($\hat{K}_\mu$ and $\hat{D}$ denote the generators of SCT 
and scale transformations, respectively) proportional to the extension 
parameter $N$, 
which characterizes the unitary irreducible 
representations of the pseudo-extended conformal group 
$\TG=SO(4,2)\tilde{\otimes}U(1)$, i.e.,  
a trivial central extension of $SO(4,2)$ by $U(1)$ with generating function 
$\eta_N(g)=N\delta$ linear in the dilatation coordinate $\delta$. 
That is, time is a quantum observable 
subject to uncertainty relations in conformally invariant quantum mechanics on 
Minkowski space and this fact extends covariance rules to the quantum domain.  
As a consequence of this, conformal wave functions 
$\psi^{(N)}(\vec{x},t)$ have 
support on the whole space-time. If we  forced the functions $\psi^{(N)}$ to 
evolve in time according to the Klein-Gordon-like equation 
\be
\hat{Q}\psi^{(N)}=\hat{P}_\mu \hat{P}^\mu \psi^{(N)}=0\,,\label{nullmass}
\ee
\ni  (``null square mass condition'', i.e., by selecting those 
functions annihilated by the 
Casimir operator $\hat{Q}$ of the Poincar\'e 
subgroup of $SO(4,2)$) we would find 
that the appearance of {\it quantum} terms proportional to $N$ at 
the right-hand side of the quantum commutators 
\be
\l[\hat{K}_\mu,\hat{Q}\r]= f_\mu(x,t)\hat{Q} +8N\hat{P}_\mu\label{malo}
\ee
\ni (where $f_\mu(x,t)$ are 
 some functions on the generalized Minkowski space),
terms  which do not appear at the classical level 
($N=0$), prevent the whole conformal group to be an exact symmetry of the 
massless quantum field. This way, the quantum time evolution  
itself destroys the conformal symmetry, 
leading to some sort of {\it dynamical symmetry breaking} which 
preserves the Weyl 
subgroup (Poincar\'e + dilatations). The SCT do not leave  
Eq. (\ref{nullmass}) invariant, as  can be deduced from (\ref{malo}), 
and this fact  manifests itself, at the second quantization level, 
through a {\it radiation} 
of the vacuum of a massless quantum field (``Weyl vacuum'') under the 
action of SCT. That is, from the point of view of a uniformly accelerated 
observer, the Weyl vacuum (which proves to be a coherent conformal state 
made of {\it zero modes}) radiates like a black body. 
The spectrum of the outgoing particles can be calculated exactly 
\cite{conforme} and proves to be a generalization of the Planckian one, this 
being recovered in the limit $N\rightarrow 0$. 
The temperature $T$ of this thermal 
bath is linear in the acceleration parameter $a$, more precisely 
$T=\frac{\hbar}{2\pi ck}a$ where $k$ denotes Boltzmann's constant 
and $c$ is the speed of light. This simple, but profound, relation 
between temperature and acceleration was first considered by 
Unruh \cite{Unruh}.

\section{Comments}

Several approaches to quantum theory (namely, canonical quantization, 
path integrals, geometric quantization, etc) exist and all of them 
are still rather rooted in the classical understanding of physical phenomena. 
Indeed, the reference to a classical action as a starting point seems 
to be the usual prerequisite to define a given quantum system. New 
perspectives on the approach to quantum physics can provide novel insights 
that could not be reached from classically-oriented formulations. Also, 
the present state of `uncertainty' for a satisfactory quantum theory of 
the gravitational field demands a sound revision of Quantum Mechanics.     

In this paper, we have tried to capture much of the 
GAQ flavour by making use of some 
relevant examples where symmetry determines the quantum physical system. 
Central extensions (and more general quantum deformations) of the 
underlying symmetry algebra provide the essential ingredient for 
a group-oriented approach to quantum theory. Among its main virtues are  
the inherent formulation of quantum theories without reference to a previous 
classical system, especially gauge theories, which exhibit a particular 
fibre-bundle structure. Nevertheless, the full richness and scope 
of this alternative point of view deserves further study.

\section*{Acknowledgment}

M. Calixto thanks the University of Granada for a Post-doctoral grant and the 
Department of Physics of Swansea for its hospitality.


\begin{thebibliography}{99}
\bibitem{gphases} Alfred Shapere and Frank Wilczek, {\it Geometric 
Phases in Physics}, Advanced series in mathematical physics Vol. 5, 
World Scientific (1989).
\bibitem{GAQ} V. Aldaya and J. de Azc\'arraga, J. Math. Phys.{\bf 23}, 
1297 (1982).
\bibitem{Ramirez} V. Aldaya, J. Navarro-Salas and A. Ram\'\i rez, 
Commun. Math. Phys. {\bf 121}, 541 (1989).
\bibitem{Extensiones} V. Bargmann, Ann. Math. {\bf 59}, 1 (1954).
\bibitem{Saletan} E. J. Saletan,  J. Math. Phys. {\bf 2}, 1 (1961).
\bibitem{Pseudoco} V. Aldaya and J.A. de Azc\'arraga, Int. J. Theor. Phys. 
        {\bf 24}, 141 (1985).
\bibitem{Marmo2} V. Aldaya, J. Guerrero, G. Marmo, {\it Quantization 
of a Lie Group: Higher Order Polarizations},  in "Symmetries in Science X", 
Ed. Bruno Gruber and Michael Ramek, Plenum Press New York (1998); 
{\sf physics/9710002}.
\bibitem{Jackiwtheta} R. Jackiw and C. Rebbi, Phys. Rev. Lett. 
{\bf 37}, 172 (1976),\\ 
C. Callan, R. Dashen and D. Gross, Phys. Lett. {\bf B63}, 334 (1976).
\bibitem{chorri} V. Aldaya, J. Navarro-Salas, J. Bisquert and R. Loll,
                 J. Math. Phys. {\bf 33}, 3087-3097 (1992).
\bibitem{symplin} M. Calixto, V. Aldaya and J. Guerrero, Int. J. Mod. 
Phys. {\bf A13}, {4889} (1998).
\bibitem{Itzykson} Claude Itzykson and Jean-Bernard Zuber, {\it Quantum 
Field Theory}, McGraw-Hill (1980).
\bibitem{Suga} H. Sugawara, \PR {\bf 170}, 1659 (1968).
\bibitem{GvH} V. Guillemin and S. Stenberg, {\it Symplectic techniques 
in physics}, Cambridge University Press (1991).
\bibitem{Abraham} R. Abraham and J.E. Marsden, {\it Fundations of 
Mechanics}, 2nd Ed., Benjamin/Cummings, Menlo Park, (1978).
\bibitem{Gotay} Mark J. Gotay, Hendrik B. Grundling and Gijs M. Tuynman,  
J. Nonlinear Sci. {\bf 6} 469 (1996), {\sf dg-ga/9605001}  \\ 
Mark J. Gotay,  J. Math. Phys. {\bf 40} 2107 (1999), {\sf math-ph/9809015}.
\bibitem{WB} E. Witten, \NP {\bf B373} (1992) 187;\\
I. Bakas, \PL {\bf B228} (1989) 57.
\bibitem{Pope} C.N. Pope, X. Shen and  L.J. Romans, \NP {\bf B339}, 191 (1990).
\bibitem{Moyal} J.E. Moyal, Proc. Cambridge Philos. Soc. {\bf 45}, 99 (1949).
\bibitem{Poly} A.M. Polyakov, Phys. Lett. {\bf B103}, 207 (1981). 
\bibitem{Pope1} E. Bergshoeff, C.N. Pope, L.J. Romans, E. Sezgin, X. Shen and 
K.S. Stelle, \PL {\bf B243}, 350 (1990).\\ 
E. Bergshoeff, P.S. Howe, C.N. Pope, E. Sezgin, X. Shen and 
K.S. Stelle, \NP {\bf B363}, 163 (1991).
\bibitem{Pope2} C.N. Pope, X. Shen, K.-W. Xu and 
Kajia Yuan, \NP {\bf B376}, 52 (1992).

\bibitem{Shen} X. Shen, Int. J. Mod. Phys. {\bf A7}, 6953 (1992).
\bibitem{Poly2} A.M. Polyakov, Mod. Phys. Lett. {\bf A2}, 893 (1987).
\bibitem{Nissimov} E. Nissimov, S. Pacheva and I. Vaysburd, Phys. Lett. 
{\bf B288}, 254 (1992).
\bibitem{progress1} M. Calixto, in progress.
\bibitem{progress2} M. Calixto, {\it Generalized W Algebras from Poisson 
Brackets on Coadjoint Orbits of Unitary Groups}, in progress.
\bibitem{coadjoint} V. Aldaya, J. Navarro-Salas and M. Navarro, 
J. Phys. {\bf A26}, 5391 (1993).
\bibitem{coadjoint1} A. A. Kirillov, {\it Elements of the theory of 
representations}, Springer, Berlin (1976).\\
      B. Kostant, {\it Quantization and Unitary
                         Representations},
                         in Lecture Notes in Math. {\bf 170}, 
                         Springer-Verlag, Berlin (1970).\\
J.M. Souriau, {\it Structure des systemes 
                         dynamiques},
                         Dunod, Paris (1970).
 \bibitem{coadjoint2} H. Aratyn, E. Nissimov, S. Pacheva and A.H. Zimmerman, 
Phys. Lett. {\bf B240}, 127 (1990); H. Aratyn, E. Nissimov and S. Pacheva, 
Phys. Lett. {\bf B251}, 401 (1990).
\bibitem{FracHall} V. Aldaya, M. Calixto and J. Guerrero,  
Commun. Math. Phys. {\bf 178}, 399 (1996).
\bibitem{Connes} A. Connes, {\it NonCommutative Geometry}, Academic Press,
Inc. , (1994).
\bibitem{modular} J.Guerrero, M. Calixto and V. Aldaya, {\it Modular 
invariance on the torus and Abelian Chern-Simons theory}, J. Math. Phys. 
{\bf 40} (1999), in press, {\sf hep-th/9707237}; 
Phys. At. Nucl. {\bf 61}, 1960 (1998).
\bibitem{Prange} R. E. Prange and S. M. Girvin, {\it The Quantum Hall 
Effect}, Springer-Verlag (1987)
\bibitem{Capelli} Andrea Cappelli {\it et all}, \PL {\bf B306}, 100 (1993); 
\NP {\bf B396}, 465 (1993); \NP {\bf B398}, 531 (1993); 
\PRL {\bf 72}, 1902 (1994); \NP {\bf B448}, 470 (1995); 
\NP {\bf B490}, 595 (1997).
\bibitem{Kac} V. Kac and A. Radul, Commun. Math. Phys. {\bf 157}, 429 (1993).
\bibitem{conforme} V. Aldaya, M. Calixto and J.M. Cerver\'o, 
 Commun. Math. Phys. {\bf 200}, 325 (1999).
\bibitem{empro} V. Aldaya, M. Calixto and M. Navarro, Int. J. Mod. Phys. 
{\bf A12}, 3609 (1997).
\bibitem{ym} M. Calixto and V. Aldaya, 
{\it Group approach to quantization 
of Yang-Mills theories: a cohomological origin of mass}, J. Phys. {\bf A32} 
(Math. Gen.), in press, (1999).
\bibitem{gtp} M. Calixto and V. Aldaya, {\it Gauge transformation properties 
of vector and tensor potentials revisited: a group quantization approach}, 
preprint {\sf hep-th/9903106}.
\bibitem{Jackiw} R. Jackiw, Rev. Mod. Phys. {\bf 52}, 661 (1980).
\bibitem{Fairlie} D.B. Fairlie and P. Fletcher, J. Math. Phys. 
{\bf 31}, 1088 (1990).
\bibitem{Kogan} Ian I. Kogan, Int. J. Mod. Phys. {\bf A9}, 3887 (1994). 
\bibitem{Maxwell}E. Cunnigham, Proc. R. Soc. Lond.{\bf 8} , 77 (1910).
\\ H. Bateman, Proc. London Math. Soc. {\bf 8}, 223 (1910).
\bibitem{Fronsdal} C. Fronsdal, Phys. Rev. {\bf D12}, 3819 (1975). 
\bibitem{Hill} E. L. Hill, Phys. Rev. {\bf 67}, 358 (1945).
\bibitem{Fulling} Stephen A. Fulling, 
Phys. Rev. {\bf D7}, 2850 (1973).
\bibitem{Unruh} W. G. Unruh, Phys. Rev. {\bf D14}, 870 (1976)
\bibitem{rusos} A.A. Grib and N. Sh. Urusova, 
Teoreticheskaya i Matematicheskaya 
Fizika {\bf 54}, 398 (1983). 
Translated in Institute of Precision 
Mechanics and Optics, Leningrad.

\end{thebibliography}
\end{document}